\documentclass[%
 reprint,
 amsmath,amssymb,
 aps,
showkeys]{revtex4-2}

\usepackage{braket}
\usepackage{subcaption}
\usepackage[autostyle]{csquotes}
\usepackage{braket}
\usepackage{graphicx}
\usepackage{dcolumn}
\usepackage{bm}

\usepackage{parskip}

\begin{document}

\preprint{APS/123-QED}

\title{Conditions for Time-Independence of N-level Systems under the Rotating Wave Approximation (RWA) and Dipole Selection Rules}

\author{Phoenix M. M. Paing}
\email{phoenix.paing@mail.utoronto.ca}
\author{Daniel F. V. James}%
 \email{dfvj@physics.utoronto.ca}
\affiliation{ 
Department of Physics, University of Toronto, 60 St George St., Toronto ON M5S 1A7
}%

\date{\today}

\begin{abstract}
We analyze the time-dependence of N-level systems under the Rotating Wave Approximation and dipole selection rules. Such systems can be solved straightforwardly if the Hamiltonian can be transformed into a time-independent form. The conditions under which a unitary transformation can be used to render time-dependent Hamiltonians into a time-independent form, thereby making the solution, are examined. After case-by-case analysis of different four and five-level systems, we conclude that systems having only one odd or even parity level achieve time-independence. In contrast, the others must satisfy a condition of laser detuning to achieve time-independence.
\end{abstract}

\keywords{Rotating Wave Approximation, N-level systems, Dipole selection rules, Hamiltonian time-independence, Rotating frame transform, System detuning conditions}                              
\maketitle

\onecolumngrid
\begin{center}
\vspace*{-1em}
\noindent\textit{
This is the author-accepted version of the manuscript that has been accepted for publication in 
\textbf{Journal of Modern Optics} (Taylor \& Francis, 2025).}
\vspace{1em}
\end{center}
\twocolumngrid

\section{\label{Intro} Introduction} 

Laser excitations and multi-level atomic systems are simple yet fundamental tools that can be applied in various fields of quantum technology. Multi-level atomic systems are used in optical atomic clocks to achieve standard timekeeping of precision up to $10^{-19}s$ \cite{andrew_ludlow, clock}. Rydberg atoms are a direct application of multilevel atomic systems and are very useful in quantum gate construction and quantum information processing \cite{saffman_rydberg}. Successful quantum gates such as the Cirac-Zoller gate \cite{cirac}, and the Molmer-Sorensen gate \cite{S_rensen_2000} are examples of the utilization of multi-level atomic systems in quantum information science. STImulated Raman Adiabatic Passage(STIRAP) uses a 3-level lambda($\lambda$) system, and it demonstrates coherent population transfer from a nearly-degenerate ground state to the other \cite{bergmann, Shore:17, bergman2}. Quantum computing techniques using qudits(quantum states with more than two qubits) strongly correlate with multi-level atomic systems \cite{Wang_2020}. The validity of the RWA on multi-level atomic systems when exposed to the environment is also an active research area \cite{open}. The multi-level systems are also interesting for their applications in nonlinear optics and photonics in characterizing various nonlinear processes such as four-wave mixing \cite{FWM, photonics, photonicsII}. A recent work from Keck et al. demonstrated how to use a control field to make the Hamiltonian time-independent under the rotating wave approximation \cite{Keck}. Beyond the RWA, counter-rotating terms can introduce significant modifications to level dynamics, such as Bloch–Siegert shifts and altered transition pathways \cite{bloch}. Recent work by Bugarth et al. and Paing explores such effects in two-level contexts, using the Jaynes-Cummings model while highlighting issues with the RWA \cite{Burgarth_2024, paing_magnus}. It can be seen that although the RWA has theoretical inconsistencies and limitations, analytical calculations beyond the RWA is challenging even for a simple two-level model of an atom. It is beyond the scope of the research to investigate non RWA contributions on higher-level systems. Therefore, it is worth revisiting what we know about the solutions of general N-level systems under RWA.\par
This research is based on a paper from Einwohner, Wong, and Garrison that studied the solvability of N-level systems using graph theory \cite{Einwohner}. There has also been work by Fujii et al. on the exact solutions of 3-level systems \cite{fujiietal, fujii}.
We ask a similar question as Einwohner, Wong, and Garrison on the condition for time-independence of various N-level systems. However, we approach it from the rotating frame and a case-by-case analysis of different systems.\par
The Hamiltonian of a general N-level system under RWA can be written as follows \cite{allen_eberly}. 
\begin{align}
    \hat{H} = \hat{H_0} + \hat{\vec{d}}\cdot \vec{E}(t)
\end{align}
\begin{align}
   \hat{H} = \sum_{n=1}^N \hbar \omega_n \ket{n}\bra{n} + \hbar(\sum_{m,n = 1}^N \frac{\Omega_{nm}}{2} e^{-i\omega_{nm}t} \ket{n}\bra{m} + c.c.)
    \label{2}
\end{align}
$H_0$ represents the Hamiltonian of an isolated N-level system with no interaction between the levels. The field of interest is a multi-mode electric field with the assumption that only one field mode close to the transition frequency gets coupled \cite{Einwohner}. Under RWA, the Hamiltonian can be represented as in equation \ref{2} where $\Omega_{nm} = \frac{1}{\hbar}\bra{n}\hat{\vec{d}}\ket{m}\cdot \vec{E}(\vec{r})$ is the Rabi frequency between the two levels and $\omega_{nm}$ is the laser frequency of the transition between two levels.\par
The dipole operator($\hat{\vec{d}}$) is a classical dipole $\hat{\vec{d}} = e\hat{\vec{r}}$ which neglects the relativistic effects of the atom. Since we are interested in atomic systems concerning ultracold atoms, the approximation is valid as the cold atoms do not move at relativistic speeds. We are also ignoring the field couplings with quadrupole, octupole, etc. Dipole selection rules allow us to reduce the number of transitions. As a result, the topology of the graphs representing various systems becomes non-trivial and of interest. Therefore, we look for conditions for complete time-independence of different topologies governing the multi-level systems. Some applications go beyond the dipole approximation in ion-trapped quantum computing \cite{H_S_Freedhoff_1989, James_1998}. Such schemes using quadrupole and octupole transitions are beyond the scope of this research.\par
The eigenstates (orbitals) of atoms are eigenstates of the parity operator. Since the position operator($\hat{\vec{r}}$) is odd, the matrix element $\bra{n}\hat{\vec{d}}\ket{m}$ is zero unless the two levels have opposite parities. Therefore, the Hamiltonian of the system depends on how many even-parity levels the system has. This paper analyzes every 4-level system while extending the analysis to induce the conditions for time-independence of a general N-level system.

\section{Brief Review on the Rotating Frame Transforms}
Solving the Schrodinger equation of Hamiltonians with explicit time dependence is a challenging task. There are a few approaches. The simplest approach of all is transforming the time-dependent Hamiltonian into a time-independent form using a unitary matrix and its conjugate as follows.
\begin{equation}
    \hat{H'} = \hat{U}^\dagger \hat{H}\hat{U} -  i\hbar \hat{U}^\dag \frac{\partial \hat{U}}{\partial t}
    \label{H_transform}
\end{equation}
If the transformed Hamiltonian is time-independent, it can be diagonalized for exact solutions. However, in this paper, we are only interested in the question of which systems can be rendered into a time-independent form instead of what the solutions are. \par
The unitary matrix used for the rotation is a diagonal unitary(equation \ref{U}).
\begin{equation}
    U = \sum_{n=1}^N e^{-i\bar{\omega}_nt}\ket{n}\bra{n}
    \label{U}
\end{equation}
The chosen unitary matrix has a few advantages. Since the phases cancel when multiplied with the conjugate, the second term of equation \ref{H_transform} has no time-dependence, and we only have to worry about $\hat{U}^\dagger \hat{H}\hat{U}$. However, a diagonal unitary has four degrees of freedom, and we will see that the degrees of freedom vs. the number of transitions determines whether a system can be transformed into a time-independent form or not. \par
Having introduced the rotating picture and the general method of the analysis, we are ready to answer the question of which systems are time-independent.

\section{Unconditionally Time-Independent Systems}
\subsection{3+1 systems}
\begin{figure}[h]
\centering
    \begin{subfigure}[b]{0.2\textwidth}
        \centering
        \includegraphics[width = 1\linewidth, height = 3cm]{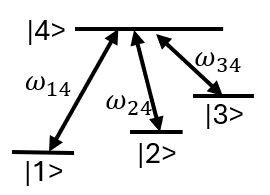}
        \caption{M system}
        \label{M}
    \end{subfigure}
    \begin{subfigure}[b]{0.2\textwidth}
        \centering
        \includegraphics[width=1\textwidth, height = 3cm]{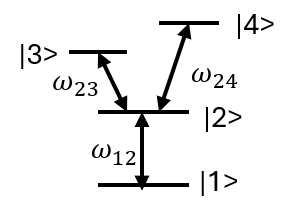}
        \caption{Y system}
        \label{Y}
    \end{subfigure}
    \hspace{0.05\textwidth}
    \begin{subfigure}[b]{0.2\textwidth}
        \centering
        \includegraphics[width=1\textwidth, height = 3cm]{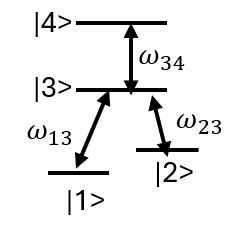}
        \caption{$\lambda$ system}
        \label{lambda}
    \end{subfigure}
    \hspace{0.05\textwidth}
    \begin{subfigure}[b]{0.2\textwidth}
        \centering
        \includegraphics[width=1\textwidth, height = 3cm]{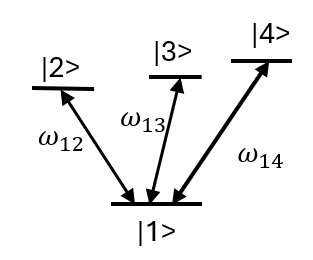}
        \caption{W system}
        \label{W}
    \end{subfigure}
    \caption{All possible diagrams of a 3+1 four-level systems. The names were chosen by the author according to symbol resemblance.}
    \label{3+1}
\end{figure}

3+1 systems are systems where three levels are even and one is odd, or three are odd and one is even. There are four different 3+1 systems, as shown in figure \ref{3+1}.
Although the Hamiltonians of the four 3+1 systems are different, they follow a pattern after transformation. We will only show the step-by-step approach to solving the $\lambda$ system, and the others can be solved using the same method.  \par
The Hamiltonian of the $\lambda$ system in matrix form is as follows.
\begin{equation*}
    \hat{H}_\lambda = \hbar\begin{pmatrix}
        \omega_1 & 0 & \frac{\Omega_{13}}{2}e^{-i\omega_{13}t} & 0 \\
        0 & \omega_2 & \frac{\Omega_{23}}{2}e^{-i\omega_{23}t} & 0\\
        \frac{\Omega_{13}^*}{2}e^{i\omega_{13}t} & \frac{\Omega_{23}^*}{2}e^{i\omega_{23}t} & \omega_3 & \frac{\Omega_{34}}{2}e^{-i\omega_{34}t} \\
        0 & 0 & \frac{\Omega_{34}^*}{2}e^{i\omega_{34}t} & \omega_4
    \end{pmatrix}
    \label{H_lambda}
\end{equation*}
Acting on the diagonal unitary and setting the time-evolving phases to zero, we get a system of equations to solve as follows.
\begin{eqnarray*}
    && \bar{\omega}_1 + \omega_{13} - \bar{\omega}_3 = 0 \\
   &&\bar{\omega}_2 + \omega_{23} - \bar{\omega}_3 = 0 \\
   &&\bar{\omega}_3 + \omega_{34} - \bar{\omega}_4 = 0 
\end{eqnarray*}
Since $\bar{\omega}_n$ are the degrees of freedom of the unitary, there exists one more $\bar{\omega}_n$ than the transition frequencies($\omega_{mn}$). Following the same analysis on M, Y, and W systems, one sees that there also exists one more degree of freedom than the transition frequency in each 3+1 system. With the freedom to choose an extra constraint for the unitary, we conclude that every 3+1 system can be transformed into a time-independent Hamiltonian without the need for any detuning conditions.
\subsection{(N-1)+1 level systems}
\begin{figure}[h]
\centering
\includegraphics[width = 0.4\linewidth, height = 3.5cm]{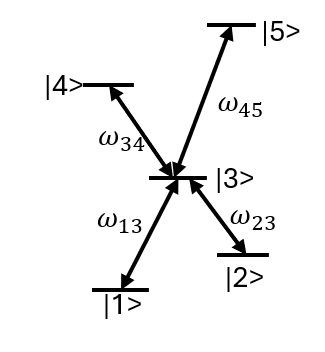}
\caption{This figure is the schematics of a one-level extension to the 3+1 $\lambda$ system.}
\label{lambda_ext}
\end{figure}

4+1 systems are just one-level extensions to the 3+1 systems. For example, let's look at a 5-level system where the third level has an opposite parity with all the other levels(figure \ref{lambda_ext}). This system is merely a one-level extension of the 3+1 $\lambda$ system.
The following is the set of equations to be solved for the time-independence of 4+1 $\lambda$ system.
\begin{eqnarray*}
    && \bar{\omega}_1 + \omega_{13} - \bar{\omega}_3 = 0 \\
   &&\bar{\omega}_2 + \omega_{23} - \bar{\omega}_3 = 0 \\
   &&\bar{\omega}_3 + \omega_{34} - \bar{\omega}_4 = 0 \\
   &&\bar{\omega}_3 + \omega_{35} - \bar{\omega}_5 = 0 
\end{eqnarray*}
It is seen that the first three equations are identical to the equations of the 3+1 $\lambda$ system. It can be shown that all the other $4+1$ systems are merely a one-level extension to the respective $3+1$ systems. This claim can be extended to N-level systems as follows: (N-1)+1 level systems are merely (N-4) level extensions to 3+1 systems. In other words, there exist (N-4) more different levels for (N-1)+1 systems, but in such systems, the degrees of freedom of the unitary are always one above the number of transitions. Therefore, we conclude that all (N-1)+1 systems can unconditionally be transformed into a time-independent Hamiltonian.

\section{Conditionally Time-Independent Systems}

We have seen in the previous section that the condition for time-independence depends on the degrees of freedom of the unitary matrix and the number of transitions of a system. For an N-level system with n number of even levels, the diagonal unitary possesses N degrees of freedom, and the number of transitions is the product of the number of even and odd levels. One would think that the condition for time-independence would be $N\leq N(N-n)$, but we will see that the system requires a detuning condition for complete time-independence when the degrees of freedom are the same as the number of transitions.
\subsection{2+2 systems}
\begin{figure}[h]
\centering
    \begin{subfigure}[b]{0.2\textwidth}
        \centering
        \includegraphics[width = 1\linewidth, height = 3cm]{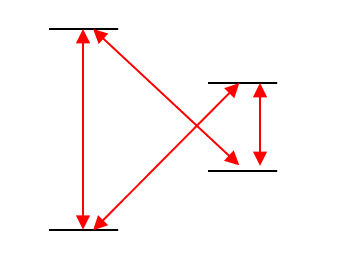}
        \caption{Hourglass system}
        \label{Hourglass}
    \end{subfigure}
    \hspace{0.03\textwidth}
    \begin{subfigure}[b]{0.2\textwidth}
        \centering
        \includegraphics[width=1\textwidth, height = 3cm]{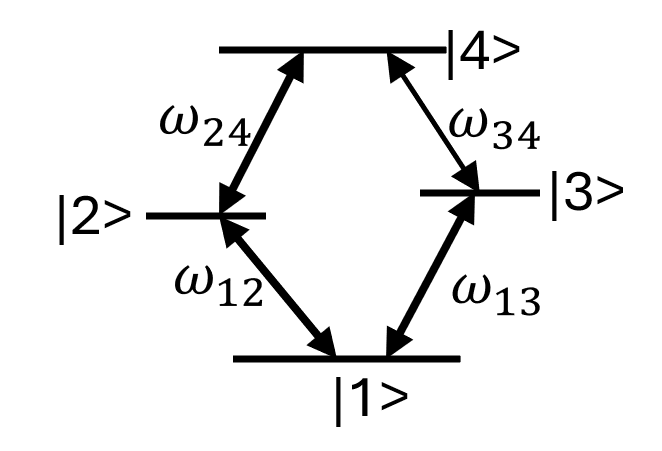}
        \caption{Diamond system}
        \label{Diamond}
    \end{subfigure}
    \hspace{0.03\textwidth}
    \begin{subfigure}[b]{0.2\textwidth}
        \centering
        \includegraphics[width=1\textwidth, height = 3cm]{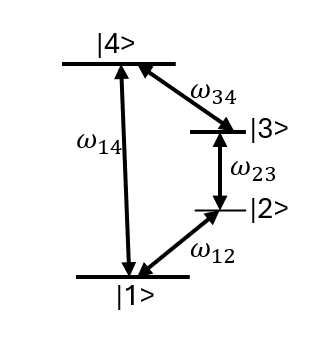}
        \caption{Trapezium system}
        \label{Trapezium}
    \end{subfigure}
    \caption{All possible diagrams of a 2+2 four-level systems.}
    \label{2+2}
\end{figure}

There are three kinds of 2+2 systems, and the schematics are shown in figure \ref{2+2}. 
The Hamiltonian of the diamond system is as follows.
\begin{eqnarray*}
    \hat{H}_D = \hbar\begin{pmatrix}
        \omega_1 & \frac{\Omega_{12}}{2}e^{-i\omega_{12}t} & \frac{\Omega_{13}}{2}e^{-i\omega_{13}t} & 0 \\
        \frac{\Omega_{12}^*}{2}e^{i\omega_{12}t} & \omega_2 & 0 & \frac{\Omega_{24}}{2}e^{-i\omega_{24}t}\\
        \frac{\Omega_{13}^*}{2}e^{i\omega_{13}t} & 0 & \omega_3 & \frac{\Omega_{34}}{2}e^{-i\omega_{34}t} \\
        0 & \frac{\Omega_{24}^*}{2}e^{i\omega_{24}t} & \frac{\Omega_{34}^*}{2}e^{i\omega_{34}t} & \omega_4
    \end{pmatrix}
\end{eqnarray*}
Acting with the unitaries and setting the time-evolving phases to zero, we get the following set of equations.
\begin{eqnarray*}
    \bar{\omega}_1 + \omega_{12} - \bar{\omega_2} = 0\\ 
    \bar{\omega}_1 + \omega_{13} - \bar{\omega_3} = 0\\
    \bar{\omega}_2 + \omega_{23} - \bar{\omega_4} = 0\\ 
    \bar{\omega}_3 + \omega_{34} - \bar{\omega_4} = 0
    \label{D}
\end{eqnarray*}
There are four degrees of freedom and four transition frequencies, so the system has the potential to be time-independent. When attempted, we encounter a condition of $\omega_{13} + \omega_{34} = \omega_{12} + \omega_{23}$. The system can be transformed into a time-independent form if the condition is met. We define the system detuning of the diamond system as $\Delta_D = \omega_{13} + \omega_{34} -\omega_{12} - \omega_{23}$, and the Hamiltonian is time-independent if the lasers are tuned in a way that the system detuning is zero.\par
The hourglass and the trapezium system can also be transformed into complete time-independence if a detuning condition is met. The detuning depends on the topology of the system, and different systems have different system detuning conditions. This paper will not go into detail about the detuning of each system. Considering the simplicity of 4-level matrix mechanics, we have found the detunings of the hourglass and trapezium systems as $\Delta_T = \omega_{14} - (\omega_{12} + \omega_{23} + \omega_{34})$ and $\Delta_H = \omega_{13} + \omega_{24} - (\omega_{23} + \omega_{14})$ respectively.\par
One final remark about the detuning is that Einwohner, Wong, and Garrison approached the same problem with graph theory, and they claimed that a system is solvable if the way to go around the graph is unchanged \cite{Einwohner}. For the diamond system, the detuning that we define is merely the difference between the way up from the left and the way up from the right. This pattern reoccurs for other 2+2 systems as well and therefore, our results agree with previous work.

\subsection{Higher Level Systems}
We have seen that (N-1)+1 level systems are time-independent since they meet the condition of time-independence: the degrees of freedom must be less than the number of transitions. For an N-level system with one even level, the degree of freedom is N, and the number of transitions is n(N-n). The only possible way of $N < n(N-n)$ is if $n=1$. Otherwise, we will have more equations than the degree of freedom of the unitary, and the system will remain time-dependent.\par
It is also evident that the higher the number of transitions is, the more equations are left unsolved, and the transformed Hamiltonian will have leftover detuning-dependent phases. Higher-level systems will have more detuning conditions upon the first rotation. We can use the same form of unitary (equation \ref{U}) to further rotate $\hat{H}'$, and we see that $\hat{H}'$ can be reduced into only one detuning-dependent term eventually. In other words, the Hamiltonian of every system except the unconditionally time-independent ones, can be reduced upon multiple rotations into the Hamiltonian with only one detuning-dependent phase. It is impossible to further reduce the time dependence of the Hamiltonian from one detuning-dependent phase. The intuitive justification is that when one of the entries of the Hamiltonian and its complex conjugate are the only time-dependent terms, there exists no other frame that can remove the time-dependence.

\section{Discussion and Conclusion}
In this paper, we have revisited the conditions for time-independece of N-level systems under the RWA. Three-level $\lambda$ systems are used in STImulated Raman Adiabatic Passage(STIRAP) for information transfer between levels \cite{bergmann}. Moreover, STIRAP can transfer the population from the ground state ($\ket{g_1}$ to a nearly-degenerate ground state with higher energy ($\ket{g_2}$) without coherence. A universal X-gate can be constructed if the reverse operation that transfers the population from $\ket{g_2}$ to $\ket{g_1}$ is implemented. We are interested in whether we can utilize higher-level systems using adiabatic transitions to construct a universal quantum logic gate. Since $(N-1)+1$ systems are unconditionally time-independent under RWA, future work will highlight will focus on solving such systems for population and coherence dynamics to search for experimental applications. \par
Moreover, the search for detuning conditions that will make a system completely time-independent remains an open area of research. We have done case by case analysis up to five-level systems, and it appears that all systems with detuning conditions can be reduced into Hamiltonians containing one detuning-dependent time-evolving phase. However, we lack concrete mathematical proof to support this claim and the graph theory seems to be the approach. \par
Nevertheless, we have seen the unconditionally time-independent systems under RWA, and the solutions should be explored, as it may be useful in constructing a quantum universal logic gate. The detuning conditions for the other systems are worth exploring, as we claim that every N-level system under the RWA is time-independent if the lasers are detuned according to the detuning conditions that depend on the topology of the system.\par
The present analysis focuses on conditions in which a multilevel atomic system can be transformed into a time-independent form under the RWA. Going beyond RWA produces additional frequency shifts, such as the Bloch-Siegert shift, which can alter the detuning conditions identified here. In some topologies, the introduction of such counter-rotating terms may prevent the complete removal of time-dependence. \par
Extending the present classification to include such effects lies beyond the scope of this article because the calculations become significantly more involved even for a simple two-level system. In our recent work, we applied the Magnus expansion to the Jaynes–Cummings model without the RWA and found that the second-order terms already yield beyond the RWA phenomena, predicting not only the Bloch–Siegert shift but also an atom-induced field-squeezing interaction \cite{paing_magnus}. The fact that such a nontrivial structure emerges in the minimal two-level case shows how challenging a full general N-level treatment would be. \par
Nevertheless, recent studies such as Bugarth et al. \cite{Burgarth_2024} and our Magnus-based analysis suggest that effective Hamiltonian and perturbative approaches could be adapted to general N-level systems to quantify beyond-RWA corrections to the time-independence criteria established here. Pursuing this extension would connect the present classification directly to experimentally observable beyond-RWA phenomena and remains an interesting challenge for future work.
\par
In this research, we have reviewed N-level systems under the RWA by a case-by-case analysis in the rotating frame. Although this approach is not the only way to solve a time-dependent Hamiltonian, it is the simplest and analytic approach to see which systems have a Hamiltonian that can be exponentiated for time-evolution.

\begin{acknowledgements}
    We would like to thank Afshin Besharat and the other participants of the Toronto Ultra Cold Atoms Network (TUCAN) conference for their comments and discussions during the conference. The comments were pivotal in the progress of the research.
\end{acknowledgements}

\section*{Disclosure of Interest}
The authors report that there are no competing interests to declare. 

\bibliographystyle{apsrev4-2} 
\bibliography{Nlvl_RWA}


\end{document}